\documentclass[reprint,amsmath,amssymb,aps,prb]{revtex4-2}
\usepackage{graphicx}
\usepackage{dcolumn}
\usepackage{bm}
\usepackage{adjustbox}
\usepackage{hyperref}
\usepackage{multirow}
\usepackage{xcolor}

\begin{document}

\preprint{APS/123-QED}

\title{Topological Hall effect in ferromagnetic Weyl semimetal Mn$_5$Ge$_3$ originating in competing dipolar interaction and magnetocrystalline anisotropy}

\author{Achintya Low}
\author{Tushar Kanti Bhowmik}
\author{Susanta Ghosh}
\author{Susmita Changdar}
\author{Setti Thirupathaiah}%
 \email{setti@bose.res.in}
\affiliation{%
 Department of Condensed Matter Physics and Material Sciences, S. N. Bose National Centre for Basic Sciences, Kolkata-700106
}%


\begin{abstract}
We report the anomalous and topological Hall effect of the ferromagnetic Weyl semimetal Mn$_5$Ge$_3$. We observe a significant anisotropic anomalous Hall effect (AHE) due to nonzero Berry curvature in the momentum space, such that the anomalous Hall conductivity (AHC) is 965 S/cm for the $xy$-plane and 233 S/cm for the $zx$-plane of the single crystal. The band structure calculations predict several Weyl and nodal points span across the momentum space, gapped out under the spin-orbit coupling effect, leading to significant $k$-space Berry curvature and large AHC. Experimentally, we also demonstrate a sizeable topological Hall effect that is originated by the non-coplanar chiral spin structure due to the competition between the out-of-plane uniaxial magnetocrystalline anisotropy and the dipole-dipole interaction between two Mn sublattices. This study hints at the importance of dipole-dipole interactions in producing the skyrmion lattice in Mn$_5$Ge$_3$.
\end{abstract}

\keywords{Suggested keywords}
\maketitle


\section{Introduction}

Triggered by the discovery of the Dirac semimetallic phase in graphene~\cite{Novoselov2005}, researchers have identified different topological materials, including the Weyl semimetals (WSMs)~\cite{Lv2015,Xu2015,Yang2015} and nodal-line semimetals (NLSs)~\cite{Bzdusek2016,Yang2018,Shukla2021}, which have bulk quasiparticle excitations following the high energy physics~\cite{Neto2009,Hasan2010}. In a WSM, the valence and conduction bands intersect linearly at a discrete (nodal) point in momentum space. On the other hand, in NLSs, no discrete nodal points exist; rather, a nodal line or loop is present in the momentum space. Weyl semimetallic phase can be found in systems with broken inversion symmetry (IS) or time-reversal symmetry (TRS). Some examples of IS broken WSM are TaAs~\cite{Lv2015}, NbP~\cite{Souma2016}, WTe$_2$~\cite{Li2017}, MoTe$_2$\cite{Deng2016}. On the other hand, the TRS is broken in magnetic Weyl semimetals. Few examples of magnetic WSMs are Co$_3$Sn$_2$S$_2$~\cite{Belopolski2021}, GdPtBi~\cite{Hirschberger2016}, Mn$_3$X (X=Sn,Ge)~\cite{Kuroda2017,Yang2017,Changdar2023}, Fe$_3$Sn$_2$~\cite{Ren2022}. The most intriguing features of WSM and NLS are the presence of net Berry curvature in the $k$-space~\cite{Hasan2010,Bansil2016}, which acts as a fictitious magnetic field on the charge carriers, leading to a large intrinsic anomalous Hall effect (AHE)~\cite{Nagaosa2010}. 

Unlike the intrinsic AHE, which is observed due to the presence of Berry curvature in the $k$-space, the topological Hall effect is the manifestation of real-space Berry curvature acquired by the conducting electrons while passing through the nontrivial chiral spin structures~\cite{Lee2009,Neubauer2009}, protected by the topological charge (Q)~\cite{Nagaosa2013,Denisov2018}. These topologically protected nontrivial spin structures are called the skyrmions, stabilized by competition among various magnetic interactions such as the Dzyaloshinskii-Moriya interactions (DMI)~\cite{Dzyaloshinskii1958,Moriya1960,Dzyaloshinskii1964,Roessler2006,Kanazawa2015}, uniaxial magnetocrystalline anisotropy~\cite{Yu2014,Duan2015,Hou2018,Preissinger2021,Purwar2023}, frustrated triangular lattice~\cite{Okubo2012,Goebel2017,Kurumaji2019,Low2022},  chiral domain-wall-induced skyrmion lattice~\cite{Gudnason2014,Yasuda2017,Cheng2019,Nagase2021,Yang2021}, and dipolar interactions~\cite{Yu2012,Kwon2012,Nagaosa2013,Heigl2021,Hassan2024}. The topological Hall effect in the noncentrosymmetric systems is unambiguously understood in terms of the competition between ferromagnetic long range exchange interactions and the DMI. On the other hand,  in the case of centrosymmetric systems, the mechanism of the topological Hall effect is a bit of a complex phenomenon involving the competition between the ferromagnetic exchange interactions and the uniaxial magnetocrystalline anisotropy, dipole-dipole interaction, local DM interactions or all interactions together. While many experimental reports demonstrated the magnetocrystalline anisotropy originated topological Hall effect~\cite{Yu2012,Yu2014,Duan2015,Hou2018,Preissinger2021,Purwar2023}, to our knowledge, no experimental study clearly evaluated the dipolar interaction induced topological Hall effect in single crystals. However, many studies demonstrated the dipolar interaction induced skyrmion lattice or topological Hall effect in multilayered or low-dimensional systems~\cite{Grundy1977,Abanov1998,Nagaosa2013,Heigl2021,Hassan2024}. Usually, the dipole-dipole interaction strength is considered small and ignored in most of the bulk systems~\cite{Wynn1997}.


\begin{figure}[t]
\includegraphics[width=\linewidth]{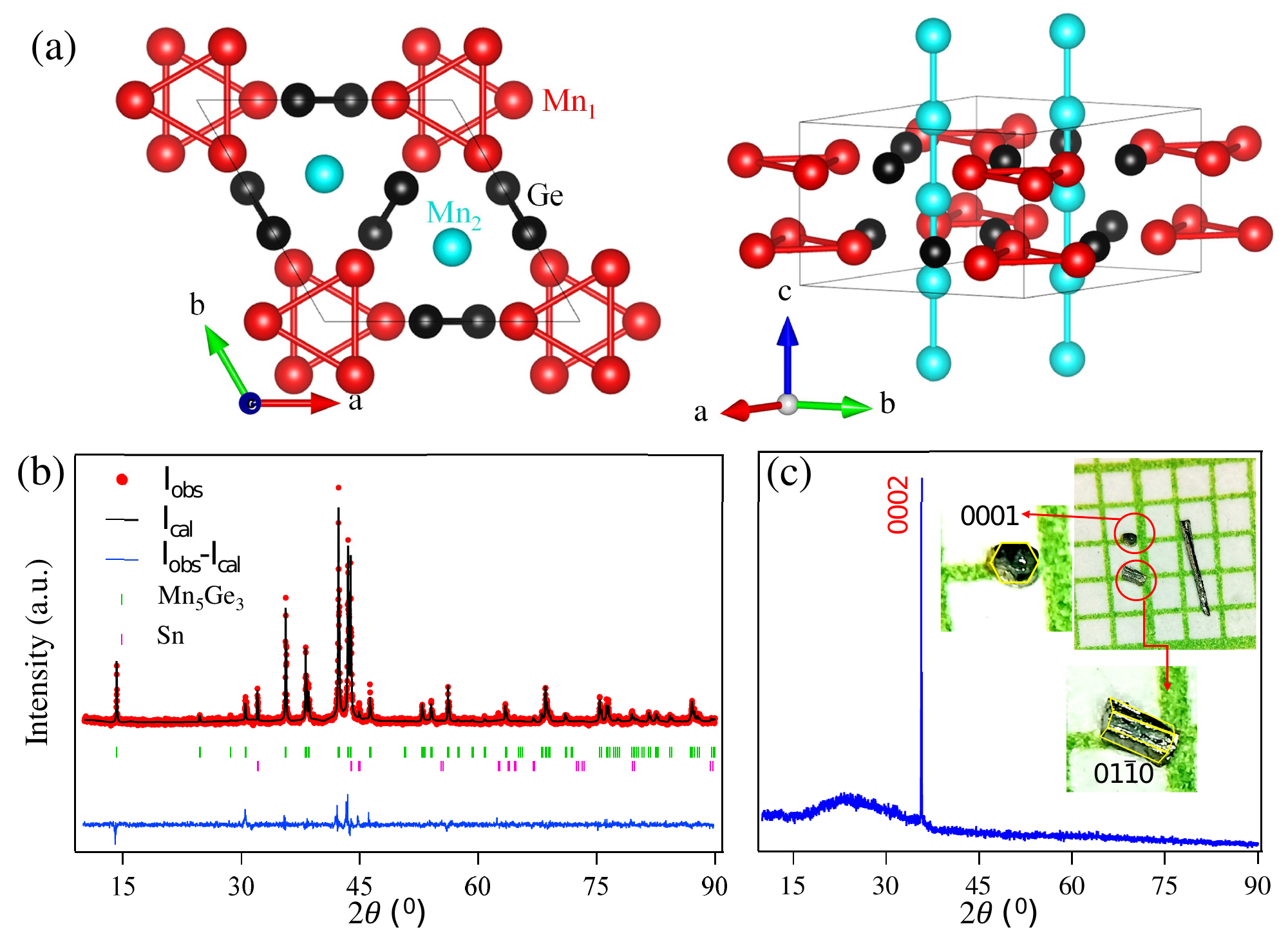}
\caption{(a) (Left) Mn$_5$Ge$_3$ crystal structure projected onto the  $ab$ plane. (Right) Hexagonal primitive unit cell of Mn$_5$Ge$_3$. (b) Powder XRD pattern of crushed Mn$_5$Ge$_3$ single crystals with overlapped Rietveld refinement. In (b), we find tiny Sn impurity peaks of the Sn flux, which is used for the crystal growth. (c) XRD pattern corresponding to the hexagonal $(0002)$ Bragg's plane. Insets in (c) show the optical images of hexagonal shaped Mn$_5$Ge$_3$ single crystal.}
\label{Fig1}
\end{figure}

Mn$_5$Ge$_3$ is an itinerant ferromagnet with a Curie temperature of about 298 K~\cite{Forsyth1990,Songlin2002,Tolinski2014}. Mn$_5$Ge$_3$ forms into the hexagonal crystal structure with a space group of $P6_3/mcm$~\cite{Forsyth1990}. Several systems with a similar space group of Mn$_5$Ge$_3$  show potential topological characteristics, such as the Dirac nodal ring in Ca$_3$P$_2$~\cite{Xie2015} and the Weyl nodal line in Eu$_5$Bi$_3$~\cite{Wu2019}. However, Mn$_5$Ge$_3$ is so far well-studied for its magnetic properties, like anisotropic magnetocaloric effect~\cite{Songlin2002,Tolinski2014,Maraytta2020} and the critical behavior analysis around magnetic transition temperature~\cite{Si2023,Lin2024}. While its sister compound Mn$_5$Si$_3$ was found to show a large topological Hall effect~\cite{Suergers2014}, a systematic topological Hall effect study is still due on Mn$_5$Ge$_3$. Most importantly, Mn atoms of this system are known to show two different sets of Wyckoff positions $4(d)$ and $6(g)$, creating two different sublattices with Mn$_{I}$ and Mn$_{II}$ type atoms, respectively~\cite{Kappel1973,Forsyth1990,Tolinski2014}. As a result, a robust dipole-dipole interaction strength has been observed between these two sublattices~\cite{Tawara1963}.

\begin{figure*}
\includegraphics[width=\linewidth]{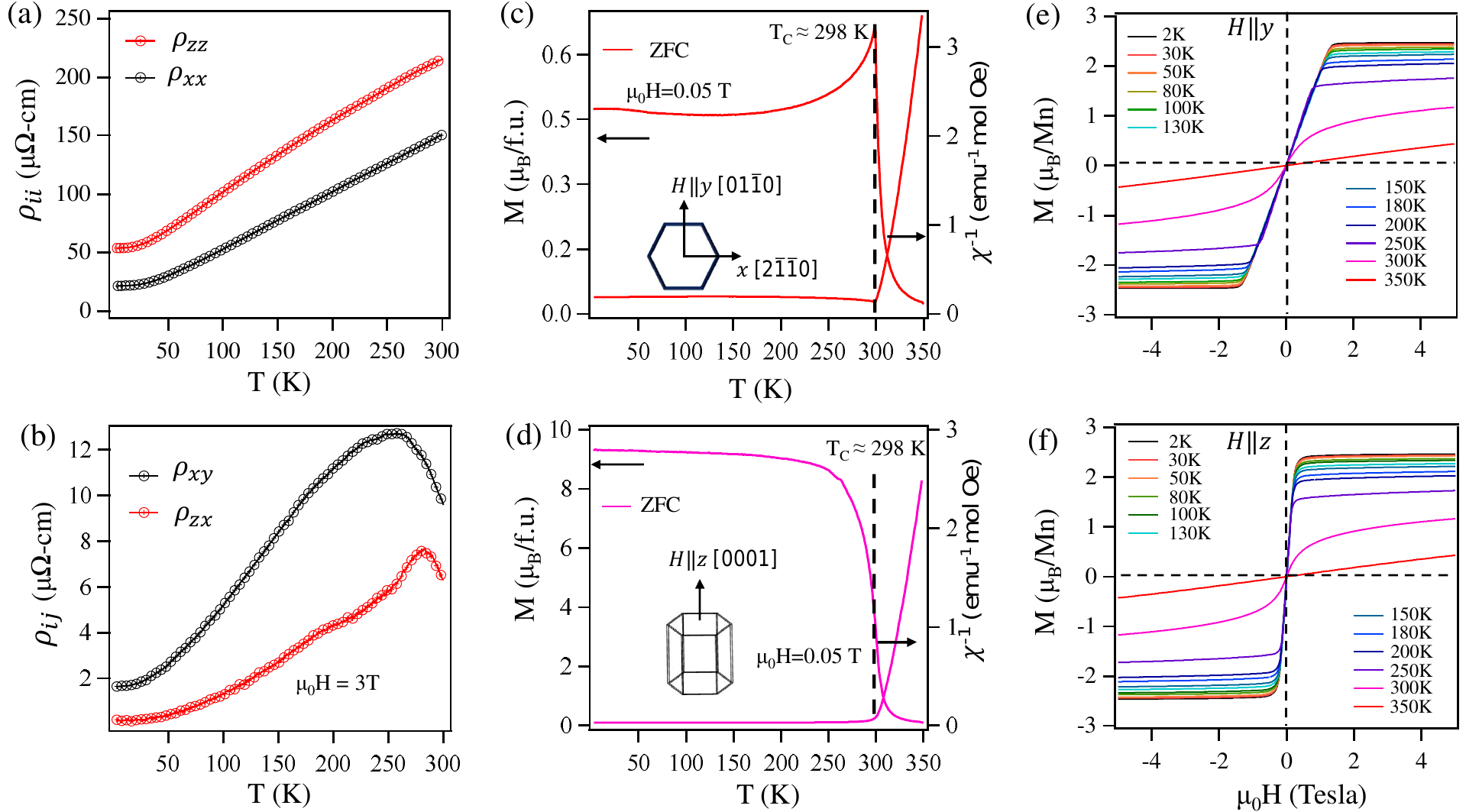}
\caption{(a) Longitudinal resistivity plotted as a function of temperature for $\rho_{xx}$ and $\rho_{zz}$ by applying current along $x$ and $z$-axis of the crystal, respectively. (b) Hall resistivity ($\rho_{xy}$ and $\rho_{zx}$) plotted as a function of temperature under 3 T of the applied magnetic field. (c) Magnetization (left axis) and inverse susceptibility (right axis) are plotted as a function of temperature [zero field cooled (ZFC)] for the fields applied along the $y$ ($[01\bar{1}0]$)-axis. (d) same as (c) except for the field applied along $z$ ($[0001]$)-axis. (e) and (f) $M(H)$ data measured at different temperatures with field applied along $y$ and $z$ directions, respectively.}
\label{Fig2}
\end{figure*}

In this study, we investigate the magnetic, anomalous, and topological Hall properties of Mn$_5$Ge$_3$ single crystal. Our studies reveal a large topological Hall effect in  Mn$_5$Ge$_3$ and a very high anomalous Hall conductivity, which is anisotropic. The $ab~initio$ calculations suggest that the spin-orbit coupling (SOC)-induced accidental gapped nodal line is a plausible cause of the large Berry curvature in this system, producing anomalous Hall conductivity. The calculations further predict several Weyl points in the momentum space. On the other hand, our experimental data suggest that the non-coplanar chiral spin structure originates the large topological Hall effect due to the competition between the out-of-plane uniaxial magnetocrystalline anisotropy and dipole-dipole interaction. Thus, this study demonstrates the importance of dipolar interactions in garnering the chiral spin structure or the skyrmion lattice in bulk systems.

\begin{figure*}[t]
\includegraphics[width=0.9\linewidth]{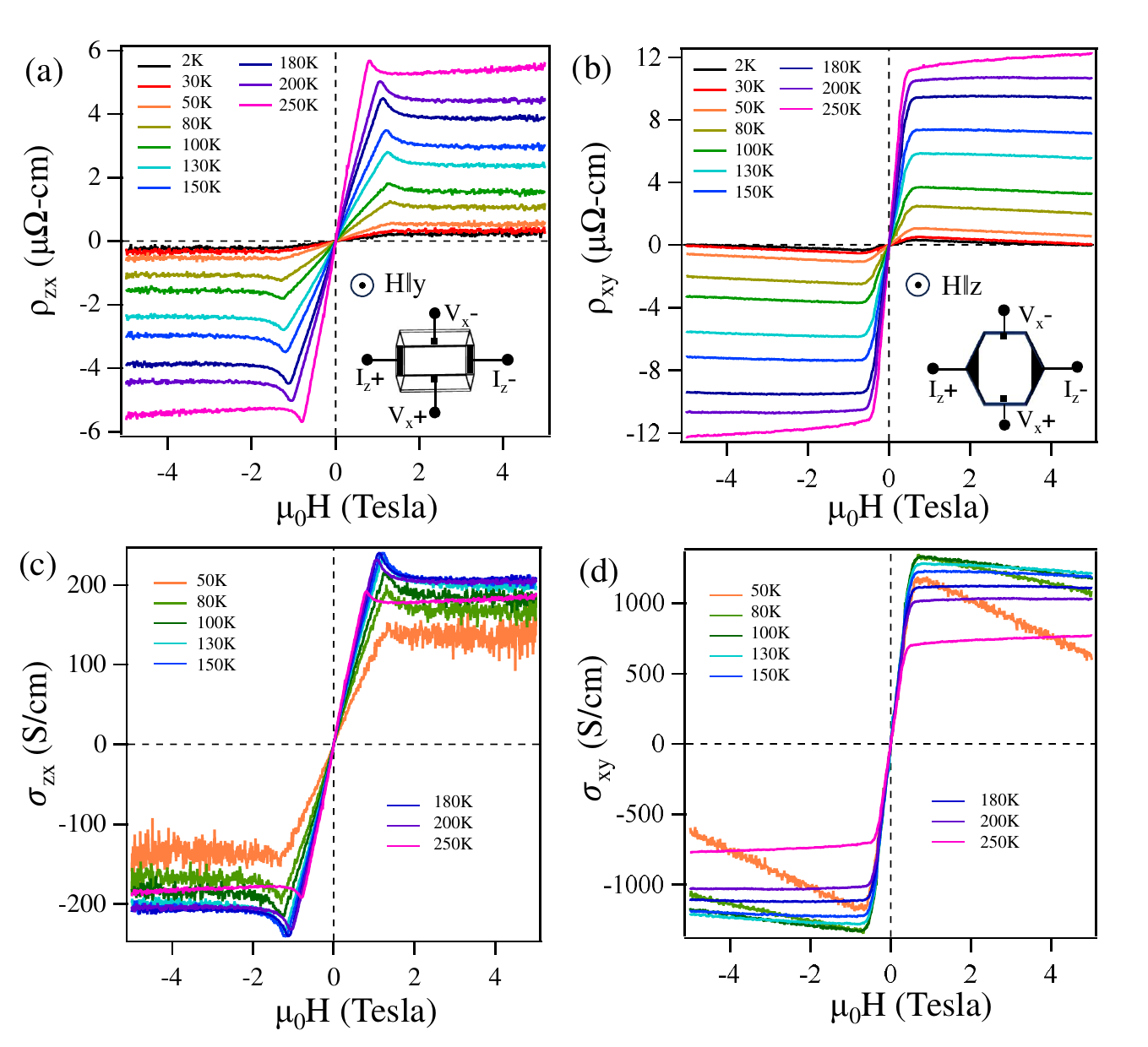}
\caption{Hall resistivity plotted as a function of the magnetic field at different temperatures for (a) $H\parallel y$ [$\rho_{zx}$] and (b) $H\parallel z$ [$\rho_{xy}$]. Schematics in the inset of (a) and (b) show the Hall resistivity measuring geometry for both directions. Hall conductivity plotted as a function of magnetic field at different temperatures are shown in (c) $\sigma_{zx}$ and (d) $\sigma_{xy}$.}
\label{Fig3}
\end{figure*}

\section{Methodology}
\subsection{Experimental Details}

Single crystals of Mn$_{5}$Ge$_3$ were grown using Sn as the flux. First, Mn powder (99.99\%, Thermo Scientific), Ge powder (99.99\%, Thermo Scientific), and Sn shots (99.998\%, Thermo Scientific) were mixed in the $1:1:5$ ratio inside an argon-filled glove box. The mixture is then placed in an alumina crucible, covered with quartz wool, and vacuum sealed inside a quartz ampoule under an argon atmosphere. Initially, the ampoule was heated up to 1000$^o$C and kept at this temperature for next 3 days, then slowly cooled down to 600$^o$C at a cooling rate of 5$^o$C/hour, and finally centrifuged at this temperature to separate single crystals from the Sn flux. This way, several rod-shaped shiny single crystals of Mn$_{5}$Ge$_3$ with a typical size of 1.5 mm$\times$0.5 mm$\times$0.5 mm were grown. Phase purity and orientation of the single crystals were checked using the X-ray diffraction (XRD) technique performed on the crushed crystals and a rod-shaped crystal using Rigaku SmartLab 9kW Cu K$_\alpha$ X-ray source. The chemical composition of the as-grown single crystals was checked using the Energy Dispersive X-ray Spectroscopy (EDS). EDS measurements suggest an actual chemical composition of Mn$_{5.05}$Ge$_{0.95}$, which is close to the nominal composition of Mn$_{5}$Ge$_3$. The linear four-probe and Hall probe connections were made using copper wires and silver paint for the electrical and magnetotransport (Hall effect) measurements. Electrical, magnetic, and magnetotransport measurements were carried out on the 9 T Physical Properties Measurement System (PPMS, Quantum Design-DynaCool) using the VSM and ETO options. Hall resistivity $\rho_{ij}$ was measured with the current applied along the $i$ direction, Hall voltage was measured along the $j$ direction,  and the field was applied perpendicular to both the $i$ and $j$ directions, where $i,j=x,y,z$. The longitudinal voltage contribution due to any misalignment of the connections was eliminated by calculating the Hall resistivity as $\frac{\rho_{H}(H)-\rho_{H}(-H)}{2}$.

\subsection{First-principles Calculations}
The electronic band structure of Mn$_5$Ge$_3$ were calculated by using density-functional theory (DFT) within the  framework of the Perdew-Burke-Ernzerhof-type generalized-gradient approximation (GGA) as implemented in the Quantum Espresso (QE) simulation package~\cite{Giannozzi2009}. The crystal structure optimization was performed using the ultrasoft pseudo-potentials~\cite{Vanderbilt1990}, with the force and the energy convergence thresholds set to 10$^{-4}$ Ry/\AA~ and 10$^{-5}$ Ry, respectively. The energy cutoff was set to 80 Ry and the charge density cutoff was set to 720 Ry for the plane wave basis, with a $k$-mesh of $10 \times 10 \times 14$ per Brillouin zone. The Marzari-Vanderbilt (mv) smearing method was employed with a smearing parameter of $\sigma=0.005$ Ry to evaluate the charge density. Spin-orbit coupling effects were treated using relativistic psuedo-potentials. To investigate the anomalous Hall conductivity (AHC), a tight-binding Hamiltonian was constructed using maximally localized Wannier functions (MLWFs) as implemented in the Wannier90 code~\cite{MOSTOFI2014}. The Berry curvature along high-symmetry directions were then calculated using the Kubo formula~\cite{Thouless1982} as implemented in Wannier90.  Subsequently, the intrinsic AHC along the [0001] and [01$\bar 1$0] directions was determined by integrating the $z$ and $y$-components of the Berry curvature over the entire Brillouin zone using the WannierTools code~\cite{Wu2017}.

\section{Results and Discussions}

Mn$_5$Ge$_3$ crystal structure consists of two types of Mn atoms. The Mn$_{I}$ atoms lie in the same plane of Ge atoms to form the triangular lattice, whereas the Mn$_{II}$ atoms form the hexagonal lattice as shown in Fig.~\ref{Fig1}(a). The powder XRD pattern of the crushed Mn$_5$Ge$_3$ single crystals is shown in Fig.~\ref{Fig1}(b). Rietveld refinement overlapped on the XRD pattern confirms the hexagonal crystal structure with a space group of $P6_3/mcm$ (no. 193) with the lattice parameters $a=b=7.201(3)~\AA$ and $c=5.039(4)~\AA$, which is in agreement with previous reports~\cite{Forsyth1990,Tolinski2014}. Fig.~\ref{Fig1}(c) presents the XRD pattern corresponding to the $(0002)$ plane of the Mn$_5$Ge$_3$ single crystal. Optical images of single crystals are shown in the insets of Fig.~\ref{Fig1}(c).


The temperature dependance of in-plane ($\rho_{xx}$) and out-of-plane ($\rho_{zz}$) electrical resistivity measured with current applied along $x$ and $z$-axis of the single crystal, respectively, are shown in Fig.~\ref{Fig2}(a).   Overall, the resistivity data suggest a metallic nature in both directions with a residual resistivity ratio [$RRR=\frac{\rho(300K)}{\rho(2K)}$] of 7 and 4 for $\rho_{xx}$ and $\rho_{zz}$, respectively. Fig.~\ref{Fig2}(b) depicts the Hall resistivity $\rho_{xy}$ and $\rho_{zx}$  as a function of temperature. Upon increasing the temperature, the Hall resistivity in both directions increases to a maximum of around 250 K and then gradually decreases. From the zero-field-cooled (ZFC) magnetization [$M(T)$] plotted as a function of temperature for both $H\parallel y$ [Fig.~\ref{Fig2}(c)] and $H\parallel z$ [Fig.~\ref{Fig2}(d)], we can see a ferromagnetic type nature from both directions with a Curie temperature of $T_C\approx 298$K. Further, the inverse susceptibility ($\chi^{-1}$) plotted as a function of temperature as shown in Figs.~\ref{Fig2}(c) and ~\ref{Fig2}(d) confirm the ferromagnetic nature of Mn$_5$Ge$_3$ as the linear curve above $T_C$ intersects the positive side of temperature axis. Importantly, after reaching a maximum $T_C$ at 298 K for $H\parallel y$, the magnetization drastically decreases with decreasing temperature down to 150 K, and then it gets saturated with further reduction in the sample temperature. Conversely, for $H\parallel z$, below $T_C$, though the magnetization increases sharply,  it only saturates below 150 K. This indicates a spin-reorientation of the Mn magnetic moments from the out-of-plane ($z$-axis) to the in-plane ($y$-axis)  above 150 K. This behavior has significance in understanding the topological Hall effect that we will be discussing later. We further measured the isothermal field-dependent magnetization [$M(H)$] for both directions as depicted in Figs.~\ref{Fig2}(e) and ~\ref{Fig2}(f). We observe no hysteresis from the $M(H)$ data,  suggesting that Mn$_5$Ge$_3$ is a soft ferromagnet. Also, the easy axis of magnetization is parallel to the $z$-axis, giving the system an out-of-plane uniaxial magnetocrystalline anisotropy. 



Next, the field-dependent Hall resistivity $\rho_{zx}$ is shown in Fig.~\ref{Fig3}(a) measured at different sample temperatures. Here, the current is along the $z$-axis,  the magnetic field is applied along the $y$-axis of the crystal, and the Hall voltage is measured along the $x$-axis. Similarly, for the Hall resistivity $\rho_{xy}$ shown in  Fig.~\ref{Fig3}(b), the current is along the $x$-axis, the field is along the $z$-axis, and the Hall voltage is measured along the $y$-axis. From the Hall resistivity data shown in Figs.~\ref{Fig3}(a) and ~\ref{Fig3}(b), we can find a large and anisotropic anomalous Hall effect from both directions. Particularly, the in-plane Hall resistivity ($\rho_{xy}$) is almost two times higher than the out-of-plane Hall resistivity ($\rho_{zx}$).   The Hall conductivity derived from the equation,  $\sigma_{zx(xy)}=-\frac{\rho_{zx(xy)}}{\rho_{zx(xy)}^2+\rho_{zz(xx)}^2}$ is plotted in Figs.~\ref{Fig3}(c) and ~\ref{Fig3}(d). Here, $\rho_{zz(xx)}$ is the longitudinal resistivity measured along $z(x)$-axis of the crystal.

\begin{figure*}[ht]
\includegraphics[width=\linewidth]{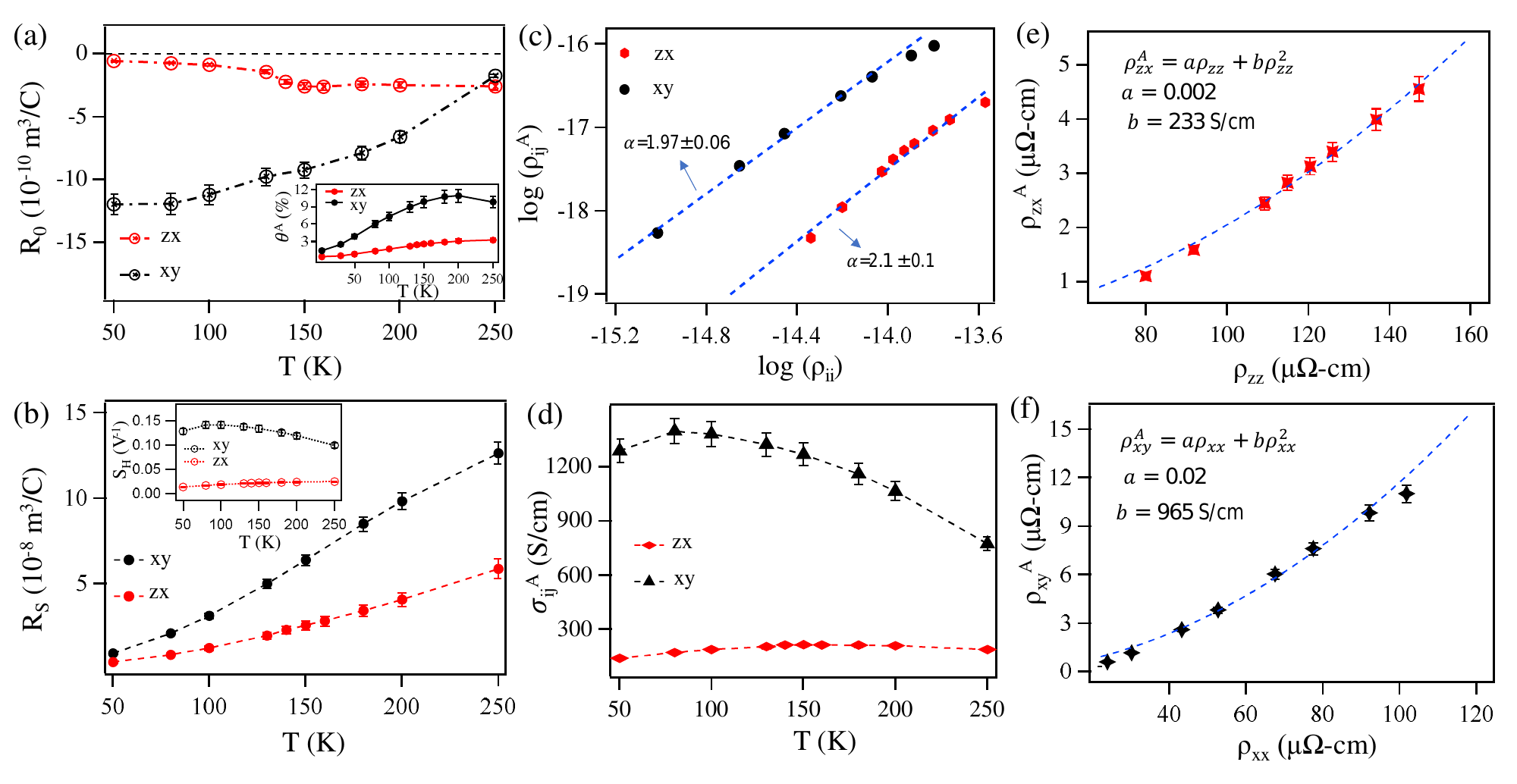}
\caption{(a) Temperature-dependent normal Hall coefficient $R_0$ plot. Inset in (a) shows anomalous Hall angle $\theta^A$ plotted a function of temperature. (b) Temperature-dependent anomalous Hall coefficient $R_S$. Inset in (b) shows anomalous Hall scaling factor. (c) log($\rho^A_{ij}$) $vs.$ log($\rho_{ii}$) plot. Dashed lines in (c) represent fitting using the relation $\rho^A_{ij}$ $ \propto$ $\rho_{ii}^{\alpha}$. (d) Temperature dependent anomalous Hall conductivity ($\sigma^A_{ij}(T)$) plot. (e) $\rho_{zx}^A$ $vs.$ $\rho_{zz}$  plot. (f) $\rho^A_{xy}$ $vs.$ $\rho_{xx}$ plot. The dashed lines in (e) and (f) represent polynomial fitting.}
\label{Fig4}
\end{figure*}

In a ferromagnet, the total Hall resistivity can be expressed by $\rho_{H}=\rho_{H}^N+\rho_{H}^A+\rho_{H}^T$~\cite{Neubauer2009,Nagaosa2010}. The first term represents normal Hall contribution. $\rho_{H}^N=\mu_0R_0H$,  where $R_0$ is the normal Hall coefficient, which can also be expressed in terms of carrier density ($n$), $R_0=\frac{1}{nq}$, $q$ is the carrier charge. The second term represents the anomalous Hall contribution, expressed by $\rho_{H}^A=\mu_0R_SM$. Finally, the last term ($\rho_{H}^T$ ) is the topological Hall resistivity contribution arising from the noncoplanar chiral spin textures. The topological Hall contribution usually vanishes at a very high magnetic field region where all the spins are fully polarized towards the applied field direction. Therefore, we fitted the Hall data at higher field regions to extract the normal and anomalous Hall contributions using the procedure as discussed in the supplemental information~\cite{supple} and in Refs.~\cite{Laha2021, Lone2024}. Fig.~\ref{Fig4}(a) depicts the normal Hall coefficient plotted as a function of temperature. As we can see from Fig.~\ref{Fig4}(a),  the Hall measurements done on the $zx$($xy$)-plane demonstrate that the dominant charge carriers are electrons. Fig.~\ref{Fig4}(b) depicts the anomalous Hall coefficient $R_S$ plotted as a function of temperature. From Fig.~\ref{Fig4}(b), we can see that the $R_S$ gradually decreases with decreasing the temperature for both directions. The anomalous Hall angle, $\theta^A=\frac{\rho_{zx/xy}}{\rho_{zz/xx}}$, which quantifies the amount of charge carriers deviating from its applied current direction, is found to be maximum \textbf{($10.8\%$)} for the $xy$-plane [see the inset of Fig.~\ref{Fig4}(a)].

 The anomalous Hall scaling factor, defined as $S_H=\frac{\mu_0R_S}{\rho_{ii}^2}$ is almost temperature independent and lies within the range of \textbf{0.01-0.14}, which is a typical range of any known ferromagnetic metal\cite{Nagaosa2010}. We further checked the scaling behavior of anomalous Hall resistivity using the relation $\rho_{ij}^A\propto \rho_{ii}^\alpha$ as shown in Fig.~\ref{Fig4}(c). We observe that for both the $zx$ and $xy$ planes, the exponent $\alpha$ is close to $2$, suggesting that the AHE is either of the intrinsic type originated from the $k$-space Berry curvature~\cite{Karplus1954} or the extrinsic type originated from the side-jump mechanism~\cite{Berger1970}. To pinpoint the correct mechanism involved in the anomalous Hall effect, in Fig.~\ref{Fig4}(d), we plotted the anomalous Hall conductivity (AHC, $\sigma_{ij}^A$) as a function of temperature. From  Fig.~\ref{Fig4}(d), we can see that the maximum AHC is found  $1400\:\mathrm{S/cm}$ from the $xy$ plane and $215\:\mathrm{S/cm}$ from the $zx$ plane.  Further,  we have fitted the anomalous Hall resistivity using the equation $\rho_{ij}^A=a\rho_{ii}+b\rho_{ii}^2$ as shown in Figs.~\ref{Fig4}(e) and ~\ref{Fig4}(f). Here, the first term represents the extrinsic skew-scattering contribution, and the second term represents either the extrinsic side-jump or the intrinsic Hall contribution. In the equation, $a$ is the skew-scattering coefficient, and $b$ is the intrinsic or extrinsic side-jump contribution coefficient.

From the fittings, we derive the values of $a=0.002$ and $b=233\:\mathrm{S/cm}$ for the $zx$ plane and $a=0.02$ and $b=965\:\mathrm{S/cm}$ for the $xy$ plane, which is very close to a previous report on this system~\cite{Zeng2006}. The side-jump contribution to the AHC can be estimated using the relation,  $\frac{e^2}{ha'}(\frac{\epsilon_{SO}}{E_F})$ where $\epsilon_{SO}$ is the spin-orbit coupling energy, $E_F$ is the Fermi energy, and $a'$ is the lattice constant~\cite{Onoda2006}. Usually, in ferromagnetic metals, the value of $\frac{\epsilon_{SO}}{E_F}$ is of the order $\sim 10^{-2}$ and  $\frac{e^2}{ha'}\approx5.97\times10^2~S/cm$ for an average lattice constant of the studied system, $a'=(2a+c)/3=6.48 ~\AA$. Therefore, the extrinsic anomalous Hall conductivity due to the side-jump $\frac{e^2}{ha'}(\frac{\epsilon_{SO}}{E_F})\approx5.97~S/cm$ is negligibly small.  Next, as for the extrinsic skew-scattering contribution, it mainly dominates at a very high conductivity region, $\sigma_{ii}>10^6 S/cm$ (clean limit)~\cite{Onoda2006}. Therefore, we can also neglect the skew-scattering contribution in high-temperature regions with very low conductivity ($\sigma_{ii}\approx 10^3$ S/cm). Finally, by ruling out both side-jump and skew-scattering extrinsic contributions, we conclude that the anomalous Hall conductivity observed in Mn$_5$Ge$_3$ is of the intrinsic type due to nonzero $k$-space Berry curvature.


\begin{figure*}
\includegraphics[width=\linewidth]{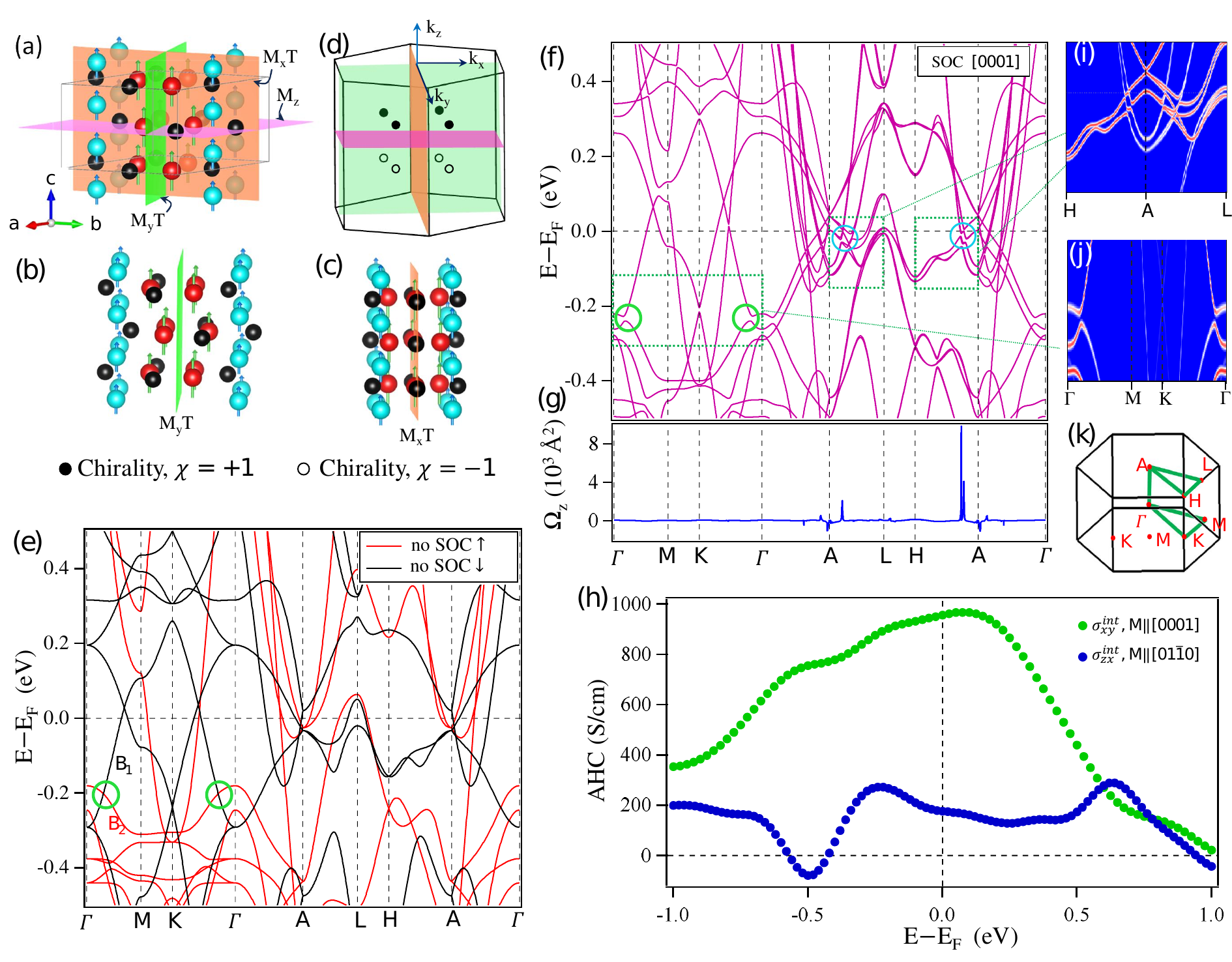}
\caption{(a)-(c) Schematically show different crystal symmetries present in Mn$_5$Ge$_3$. (d) Schematically show the presence of Weyl points in the hexagonal Brillouin zone having different chirality ($\chi$). (e) Spin-polarised electronic structure of Mn$_5$Ge$_3$ for the magnetization vector along the $[0001]$ ($z$-axis) direction, calculated without applying SOC. (f) Same as (e) but calculated with SOC effect. (g) Berry curvature ($\Omega_z$) is calculated for the magnetization vector along the $z$-axis. (h) In-plane ($\sigma^A_{xy}$) and out-of-plane ($\sigma^A_{zx}$) anomalous Hall conductivity calculated with the magnetization vector kept along the $[0001]$ ($z$) and $[01\bar{1}0]$ ($y$) directions. (i) and (j) show the zoomed-in band dispersions of (e).  (k) Schematically show the high symmetry points on the Hexagonal Brillouin zone.}
\label{Fig5}
\end{figure*}



To further confirm that the AHC contribution is from the Berry curvature, we performed the \textit{ab~initio} calculations for the electronic band structure. The calculations suggest that the two types of Mn atoms (Mn$_I$ and Mn$_{II}$) have different magnetic moments. That means the four Mn$_I$ atoms at the sites \textit{4(d)} have an average magnetic moment of $2.08~ \mu_B/Mn_I$ and the six Mn$_{II}$ atoms at the sites \textit{6(g)} have an average magnetic moment of $2.93~\mu_B/Mn_{II}$ giving an average total magnetic moment of $2.59~\mu_B/Mn$. The predicted magnetic moment is in excellent agreement with our experimental net magnetic moment of $2.46~\mu_B/Mn$.  Being the $z$-axis as the easy magnetization axis, the magnetic ground state is considered to be quantized along the $[0001]$ direction. Given the magnetic configuration of the Mn spins as shown in Fig.~\ref{Fig5}(a),  there can be a $M_z$ mirror plane. Further, there exist two mirror planes $M_x$ and $M_y$ added with half-lattice $c/2$ translational and time-reversal symmetry ($\it{T}$), $\it{T}\{M_y|\tau=c/2\}$ and $\it{T}\{M_x|\tau=c/2\}$, that are nonsymmorphic in nature. All three symmetries are crucial in forming the Weyl points in momentum space. As discussed in Ref.~\cite{Yang2017}, the mirror symmetry and the time-reversal symmetry (TRS) act on the Weyl node with a chirality $\chi$ at a momentum vector \textbf{k}($k_x, k_y, k_z$) in such a way that the mirror reflection reverses the sign of chirality ($\chi$), TRS reverses the sign of Berry curvature \textbf{($\Omega$)}, while both mirror reflection and TRS reverse the sign of momentum vector \textbf{k}.


\begin{table*}[ht]
\caption{\label{Tab1} Momentum (\textbf{$k$}), energy relative to $E_F$, Chern number, and multiplicity of the Weyl points.}
\begin{ruledtabular}
\begin{tabular}{ccccccc}
 Weyl point&$k_x$ (${\AA}^{-1}$)&$k_y$ (${\AA}^{-1}$)&$k_z$ (${\AA}^{-1}$)&$E-E_F$ (eV)& Chern number& Multiplicity\\ \hline
 W$_1$&$0$&$\pm0.16$ &$+0.33$&$-0.11$&$+1$ &$2$ \\
 W$_1$&$0$&$\pm0.16$ &$-0.33$&$-0.11$&$-1$ &$2$ \\
 W$_2$&$\pm0.07$&$\pm0.04$ &$+0.18$&$-0.08$&$+1$ &$4$ \\
 W$_2$&$\pm0.07$&$\pm0.04$ &$-0.18$&$-0.08$&$-1$ &$4$ \\
 W$_3$&$0$&$\pm0.09$ &$+0.16$&$-0.075$&$+1$ &$2$ \\
 W$_3$&$0$&$\pm0.09$ &$-0.16$&$-0.075$&$-1$ &$2$ \\
\end{tabular}
\end{ruledtabular}
\end{table*}

\begin{figure*}
\includegraphics[width=0.96\linewidth]{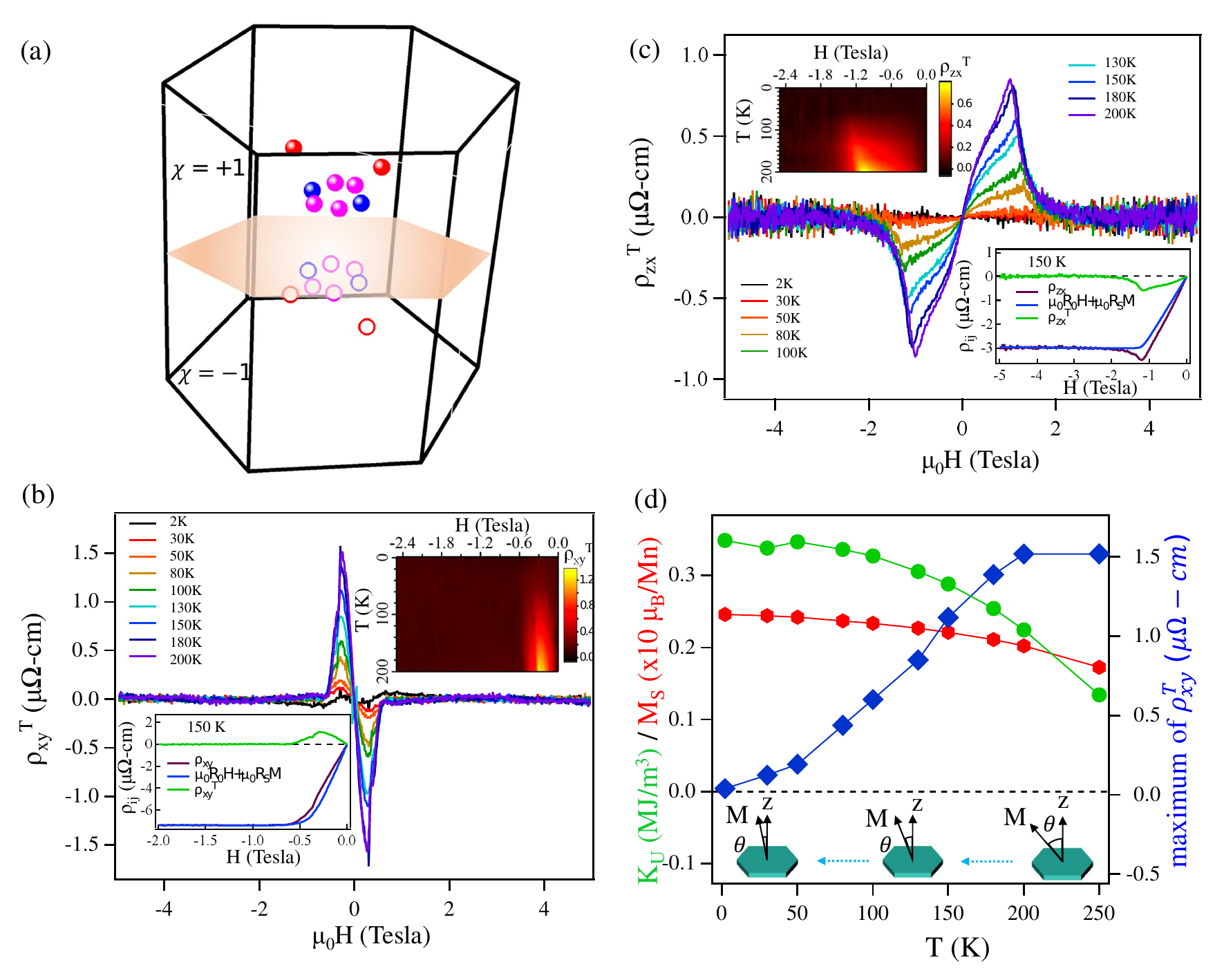}
\caption{(a) Weyl points of Mn$_5$Ge$_3$ located on the hexagonal Brillouin zone. (b) Out-of-plane ($\rho_{zx}^T$) and (c) in-plane ($\rho_{xy}^T$) topological Hall resistivity plotted as a function of the field at different sample temperatures. The top inset in (b) and (c) shows the topological Hall resistivity phase diagram. Bottom inset in (b) and (c) demonstrate the extraction of topological Hall resistivity from the total Hall resistivity measured at 150 K. (d) Show temperature-dependent uniaxial magnetocrystalline anisotropy energy density ($K_U$) and saturation magnetization ($M^z_s$) on the left-axis and topological Hall resistivity on the right-axis. Schematics at the bottom of (d) demonstrate the change of angle ($\theta$) between the easy-magnetization axis and the $z$-axis of the crystal with temperature.}
\label{Fig6}
\end{figure*}

Therefore, upon applying the symmetry operations [$M_z$, $\it{T}\{M_x|\tau=c/2\}$), and $\it{T}\{M_y|\tau=c/2\}$], we can generate a maximum of eight nonequivalent Weyl points in the $k$-space for every single Weyl point located at $\textbf{k}$,  as demonstrated in Fig.~\ref{Fig6}(a). Out of these eight Weyl points,  four are with `+$\chi$' chirality located at $(\pm{k_x}, \pm{k_y}, +k_z)$ momenta and the other four with `-$\chi$' chirality would be at $(\pm{k_x}, \pm{k_y}, -k_z)$ as tabulated in Tab.~\ref{Tab1}.

Fig.~\ref{Fig5}(e) depicts the spin-resolved electronic band structure calculated along the $k$-path shown in the figure for the magnetization vector along the $[0001]$ direction ($z$-axis) without including the spin-orbit coupling (SOC). Interestingly, at around $0.2$ eV below the Fermi level along $\Gamma-M$ and $\Gamma-K$ paths, we can see linear band crossings between the spin-up and spin-down states, forming a nodal ring around the $\Gamma$-point. In Ca$_3$P$_2$, which has a similar space group of Mn$_5$Ge$_3$, the band crossings are protected by C$_{2v}$ point group symmetry with four irreducible representatives, creating a Dirac nodal ring around the $\Gamma$ point~\cite{Xie2015}. However, in the case of Mn$_5$Ge$_3$ one has to consider the spin part as well, since it is a ferromagnetic system. As a result, C$_{2v}$ point group is replaced by C$_{2v}$ double group, which has only one irreducible representation leading to a gapped nodal ring under the SOC~\cite{Yang2018} [see Fig.~\ref{Fig5}(f)]. The same is confirmed from the orbital projected band structure calculated using the SOC [see Figs.~\ref{Fig5}(i)-(j)], where one can notice the gapped spin-up and spin-down states. In addition to this, along the paths $H-A$ and $A-L$, we see several accidental gapped nodal points that are induced due to SOC [see Fig.~\ref{Fig5}(i)].

The $z$ component of Berry curvature ($\Omega_z$) near the Fermi level is calculated along the high symmetry path as depicted in Fig.~\ref{Fig5}(g) using the relation~\cite{Xiao2010},

\begin{equation}
\Omega_{\gamma}^n(\textbf{k})=2i\hbar^2\sum_{m\neq n}\frac{\langle u_n(\textbf{k})|\hat{v}_{\alpha}|u_m(\textbf{k})\rangle\langle u_m(\textbf{k})|\hat{v}_{\beta}|u_n(\textbf{k})\rangle}{(\epsilon_n(\textbf{k})-\epsilon_m(\textbf{k}))^2}
\end{equation}

where $\hat{v}_{\alpha}=\frac{1}{\hbar}\frac{\partial\hat{H}}{\partial k_{\alpha}}$ is velocity operator, $|u_{n}(\textbf{k})\rangle$ and $\epsilon_n(\textbf{k})$ are eigenstates and eigenvalues of the Hamiltonian $\hat{H}$, respectively. Finite Berry curvature peaks along the $A-L$ and $H-A$ paths originated from the accidental gapped nodal points.

Next, the Hall conductivity is calculated at the Fermi level using the Kubo formalism~\cite{Gradhand2012},

\begin{equation}
\sigma_{\alpha\beta}=-\epsilon_{\alpha\beta\gamma}\frac{e^2}{\hbar}\sum_n\int_{BZ}\frac{d^3k}{(2\pi)^3}\Omega_{\gamma}^n(k)f_n(k)
\end{equation}

where $\epsilon_{\alpha\beta\gamma}$ is Levi-Civita tensor ($\alpha,\beta,\gamma=x,y,z$), $n$ is the band index, $\Omega_{\gamma}^n$ is the Berry curvature along the $\gamma$ axis of the momentum space, and $f_n$ is the Fermi distribution function.

The anomalous Hall conductivity for both in-plane ($\sigma_{xy}$) and out-of-plane ($\sigma_{zx}$) are plotted within the energy window of $-1$  and $+1$ eV as shown in Fig.~\ref{Fig5}(h). For the $\sigma_{xy}$ ($\sigma_{zx}$), the magnetization vector was set along the $[0001]$ ($[01\bar{1}0]$) direction.  Our calculations predict AHC values of $\sigma_{xy}=960$ S/cm and $\sigma_{zx}=200$ S/cm  near the Fermi level. The predicted AHC values agree with the experimental values of $\sigma_{xy}=965$ S/cm and $\sigma_{zx}=233$ S/cm.

Since Mn$_5$Ge$_3$ is a potential candidate for hosting the Weyl points, a search for them in the ferromagnetic $[0001]$ ground state has been conducted. We could find several Weyl nodes, with the closest ones to the Fermi level located at around 75 meV. Based on the binding energy positions, we primarily categorize three different Weyl points, W$_1$, W$_2$, and W$_3$ [see Tab~\ref{Tab1}]. However, as discussed earlier,  the Weyl point $W_1$ becomes four copies of inequivalent Weyl points located at different momenta due to various crystal symmetries. Similarly, the Weyl points $W_2$ and $W_3$ become eight and four copies of inequivalent Weyl points, respectively. The list of Weyl points, along with their momenta,  chirality, and multiplicity, is presented in Tab.~\ref{Tab1}. Fig.~\ref{Fig6}(a) schematically depicts the predicted Weyl points in the hexagonal Brillouin zone.


While fitting the field-dependent Hall resistivity data as shown in Figs.~\ref{Fig3}(a) and ~\ref{Fig3}(b),  we noticed that the normal ($\rho_{H}^N$) and anomalous Hall ($\rho_{H}^N$) contributions are not entirely reproducing the total Hall resistivity. Therefore, additional contribution from the topological Hall resistivity term is required to fit the Hall data properly. The topological Hall resistivity can be extracted by subtracting the normal and anomalous Hall resistivity from the total Hall resistivity as $\rho_{H}^T=\rho_{H}-(\rho_{H}^N+\rho_{H}^A)$. In this way, we derived the topological Hall resistivity for the $xy$ and $zx$-planes as shown in Fig.~\ref{Fig6}(b) and ~\ref{Fig6}(c), respectively. We find a maximum topological Hall resistivity above 200 K, which decreases with decreasing temperature and completely vanishes below 50 K. The top right insets of Fig.~\ref{Fig6}(b) and ~\ref{Fig6}(c) depict the H-T phase diagrams of the topological Hall resistivity for the $xy$ and $zx$ planes,  respectively. For  $xy$-plane,  the topological Hall effect is triggered in the high-temperature region within the field range of -0.5 and 0.5 T. Whereas, for $zx$-plane, the topological Hall is visible in a wider applied field range of -1.5 and 1.5 T. The topological Hall resistivity mainly originates from the non-zero scalar spin chirality [$\chi_{ijk}=(\delta \bm{S_i}\: .[\delta \bm{S_j} \times \delta \bm{S_k}$])] induced by the non-coplanar spin structure (skyrmion lattice)~\cite{Neubauer2009,Denisov2018}.

There are many ways of producing the non-coplanar spin texture (skyrmion lattice) in magnetic systems, such as Dzyaloshinskii-Moriya (DM) interaction or asymmetric exchange interaction in the inversion symmetry broken systems~\cite{Neubauer2009,Leroux2018}, Ruderman-Kittel-Kasuya-Yosida (RKKY) interaction in rare-earth-based geometrically frustrated triangular magnets~\cite{Kurumaji2019,Hirschberger2019}, or competition between uniaxial anisotropy along with the dipole-dipole interactions in centrosymmetric systems~\cite{Yu2014,Hou2018}. The first two possibilities can be excluded as our system is centrosymmetric and a 3d-transition metal-based ferromagnet. As mentioned earlier, Mn$_5$Ge$_3$ is a uniaxial ferromagnet with an easy magnetization axis along the crystal's $[0001]$ axis. Therefore, the uniaxial magnetocrystalline anisotropy energy density ($K_U$) is calculated using the formula, $K_U=\mu_0\int_{0}^{M_s}[H_y^{eff}(M)-H_z^{eff}(M)]dM$ after eliminating the geometrical demagnetization factor ($N_d$), where $H^{eff}=H-N_dM$. Here, the demagnetization factor for $H\parallel z$ is calculated to be $N_d=0.04$, and for $H\parallel y$, it is $N_d=0.48$.

Fig.~\ref{Fig6}(d) presents $K_U$ plotted as a function of temperature. From Fig.~\ref{Fig6}(d), we can notice that $K_U$ decreases monotonically with increasing the temperature, having the highest $K_U$ of 0.35 $MJ/m^3$ at 2 K. Such a monotonic decrease in $K_U$ with increasing temperature suggests that at the lowest temperature, the Mn magnetic moments quickly tend to align parallel to the $z$-axis, but as the temperature increases, the Mn magnetic moments gradually direct away from the $z$-axis. In this regard, a previous study on Mn$_5$Ge$_3$ demonstrated the presence of magnetic interaction between the Mn$_I$ [$4(d)$] and Mn$_{II}$ [$6(g)$] sublattices, leading to the dipole-dipole anisotropic energy density coefficient of $K_{dip}=(6.13\mu_1^2+3.2\mu_2^2-7.62\mu_1\mu_2)\frac{32\pi\mu_B^2}{V^2}$,  where $\mu_1=2.08\:\mu_B$ is magnetic moment of Mn$_I$ atom and $\mu_2=2.93\:\mu_B$ is magnetic moment of Mn$_{II}$ atom as per our DFT calculations. Upon substituting these $\mu_1$ and $\mu_2$ values, we calculated the dipole-dipole interaction energy density coefficient of $K_{dip}=0.04$MJ/m$^3$ which is significant and is more than $10\%$ of the uniaxial anisotropy energy coefficient ($K_U$).

The strength of dipole-dipole anisotropy energy density depends on the angle between the easy axis of magnetization and the dipole moment direction. If  $\theta$ is the angle between easy-magnetization axis and the $z$ axis, then the dipole-dipole anisotropy energy density is given by $E_{dip}=K_{dip}{\mathrm{sin}}^2\theta$ \cite{Tawara1963}. As it is evidenced from the saturation magnetization ($M^z_s$) $vs.$ temperature plot, shown in Fig.~\ref{Fig6}(d), the magnetic moments align along the $z$-axis causing the highest $M^z_s$ value at low temperature, which then decreases with increasing temperature, leading to the magnetic moments directing away from the $z$-axis. Therefore, in this scenario, the dipole-dipole anisotropic energy density would be negligible at low temperatures as $\theta\approx0^\circ$. However, $E_{dip}$ should be significantly high at higher temperatures as the magnetic moments direct away from the $z$-axis, causing larger $\theta$. Interestingly, from the maximum topological Hall resistivity ($\rho^{T}_{xy}$) $vs.$ temperature plot as shown in Fig.~\ref{Fig6}(d),  we find that the topological Hall resistivity vanishes at low temperature ($<$ 50 K), while it gradually increases with increasing temperature. The discussion of magnetic moment reorientation from the out-of-plane to in-plane with increasing temperature is in line with the $M(T)$ data shown in Figs.~\ref{Fig2}(c) and ~\ref{Fig2}(d). Therefore, the topological Hall effect observed in Mn$_5$Ge$_3$ is mainly driven by the non-coplanar spin structure originating from the competition between dipole-dipole interactions and the uniaxial magnetocrystalline anisotropy. It is worth mentioning here that the dipolar interaction-induced topological Hall effect is usually observed in low-dimensional systems~\cite{Grundy1977,Abanov1998,Nagaosa2013,Heigl2021,Hassan2024}. However, due to the peculiar magnetic interactions between Mn$_{I}$ and Mn$_{II}$ sublattices, the dipolar interactions are significant in Mn$_5$Ge$_3$.

Finally, we want to finish our discussion by mentioning a preprint that appeared during our manuscript preparation on the anomalous and topological Hall effect studies of Mn$_5$Ge$_3$ single crystal~\cite{Li2025}. The magnetotransport properties presented in Ref.~\cite{Li2025} are consistent with our observations. Most importantly, Ref.~\cite{Li2025} has reported opposite helicity for the skyrmions using a Lorentz Transmission Electron Microscope (LTEM). The opposite helicity for the skyrmions is only possible if these originate from dipole-dipole interactions~\cite{Yu2012,Kwon2012,Nagaosa2013,Heigl2021,Hassan2024}. Thus, our suggestion of dipolar interaction induced topological Hall effect in Mn$_5$Ge$_3$ is consistent with the LTEM studies in Ref. ~\cite{Li2025}.  

\section{Summary}

In summary, we have successfully grown single crystals of Mn$_5$Ge$_3$. Our studies reveal a large topological Hall effect in  Mn$_5$Ge$_3$ and a very high anomalous Hall conductivity, which are anisotropic. The density functional theory calculations suggest that the spin-orbit coupling (SOC)-induced accidental gapped nodal line produces a large Berry curvature and anomalous Hall conductivity. The noncoplanar chiral spin structure originates the significant topological Hall effect observed in this system due to the competition between the out-of-plane uniaxial magnetocrystalline anisotropy and dipole-dipole interaction between two Mn sublattices. Finally, this study unambiguously demonstrates the importance of dipolar interactions in garnering the skyrmion lattice in the bulk systems.

\begin{acknowledgments}

The authors thank the Science and Engineering Research Board (SERB), Department of Science and Technology (DST), India for the financial support (Grant No. SRG/2020/000393). This research has made use of the Technical Research Centre (TRC) Instrument Facilities of S. N. Bose National Centre for Basic Sciences, established under the TRC project of Department of Science and Technology, Govt. of India.

\end{acknowledgments}

\nocite{*}

\bibliography{MnGe.bib}

\end{document}